\documentclass[12pt]{article}
\usepackage{epsfig}
\usepackage{amsfonts}

\def\addcite#1{[???]}

\def\gray{\special{ps: 0.4 setgray}}
\def\black{\special{ps: 0.0 setgray}}

\newcommand{\draft}{
\newcount\timecount
\newcount\hours \newcount\minutes  \newcount\temp \newcount\pmhours

\hours = \time
\divide\hours by 60
\temp = \hours
\multiply\temp by 60
\minutes = \time
\advance\minutes by -\temp
\def\hour{\the\hours}
\def\minute{\ifnum\minutes<10 0\the\minutes
            \else\the\minutes\fi}
\def\clock{
\ifnum\hours=0 12:\minute\ AM
\else\ifnum\hours<12 \hour:\minute\ AM
      \else\ifnum\hours=12 12:\minute\ PM
            \else\ifnum\hours>12
                 \pmhours=\hours
                 \advance\pmhours by -12
                 \the\pmhours:\minute\ PM
                 \fi
            \fi
      \fi
\fi
}
\def\fullclock{\hour:\minute}
\begin{centering}
\gray
\special{ps: -90 rotate}
\special{ps: -5100 -5000 translate}
\font\Hugett  =cmtt12 scaled\magstep3
{\Hugett Draft: \today,\clock}
\black
\special{ps: 90 rotate}
\special{ps: 5000 -5100 translate}
\end{centering}
\vskip -1.7cm
$\phantom{a}$
} 
\catcode`\@=11 
\def\lsim{\mathrel{\mathpalette\@versim<}}
\def\gsim{\mathrel{\mathpalette\@versim>}}
\def\@versim#1#2{\vcenter{\offinterlineskip
        \ialign{$\m@th#1\hfil##\hfil$\crcr#2\crcr\sim\crcr } }}
\catcode`\@=12 

\def\nextline{\hfill\break}

\def\mycomm#1{\nextline\strut\kern-3em{\tt ====> #1}\nextline}

\def\nextline{\hfill\break}

\newcommand{\beq}{\begin{equation}}
\newcommand{\eeq}{\end{equation}}
\newcommand{\bea}{\begin{eqnarray}}
\newcommand{\eea}{\end{eqnarray}}
\newcommand{\Seff}{S_{\kern-0.1em\hbox{\small \it eff}}}
\newcommand{\Leff}{{\cal L}_{\kern-0.1em\hbox{\small \it eff}}}

\def\eqref#1{(\ref{#1})}

\def\downstrut{\vrule height 0ex depth 0.7ex width 0pt}

\begin{document}

\begin{flushright}
TAUP--2705-02\\
WIS/20/02-MAY-DPP
\end{flushright}
\vskip0.5cm
\begin{center}
{\Large\bf Scattering and Resonances in QCD$_2$}
\end{center}
\medskip
\begin{center}
{\bf Yitzhak Frishman}\footnote{\tt yitzhak.frishman@weizmann.ac.il}\\
{\em Department of Particle Physics\\
Weizmann Institute of Science\\
76100 Rehovot, Israel}\\
\vspace*{0.5cm}
and 
\\
\vspace*{0.5cm}
{\bf Marek Karliner}\footnote{\tt marek@proton.tau.ac.il}\\
{\em School of Physics and Astronomy\\
Raymond and Beverly Sackler Faculty of Exact Sciences\\
Tel Aviv University, Tel Aviv, Israel}\\
\end{center}
\begin{abstract}
Extending previous works on the spectrum of QCD$_2$, we now 
investigate the 2D analogue of meson-baryon scattering.
We use semi-classical
methods, perturbing around classical soliton solutions.
We start with the abelian case, corresponding to one flavor, and 
find that in this case the effective potential is reflectionless.
We obtain an explicit expression for the forward phase shift.
In the non-Abelian case of several flavors, the method yields a potential
which depends on the momentum of the incoming particle. In this
case there is both transmission and reflection.
In both cases no resonances appear.
As a byproduct, we derive the general conditions 
for a 2D scalar quantum field theoretical action to
yield a reflectionless effective potential when one expands
in small fluctuations about the classical solution.
\end{abstract}
\vfill
\eject

\section*{Introduction}

The semi-classical approximation has been an important tool in Quantum
Field Theory for some time now \cite{Gervais:zg}.

Typically one first finds a classical solution of the field equations,
yielding some topologically stable object (an instanton, 
a monopole, a soliton, etc.)  which cannot be
reached by perturbation theory \cite{Rajaraman}.

One then computes quantum fluctuations around such a soliton, 
yielding the quantum eigenstates,
as well as the scattering matrix for interaction
of the fluctuating fields with the soliton. 

In the context of QCD this procedure has been applied to
the Skyrme model \cite{Skyrme:vq},
which is the same universality class as the infrared limit of
large-$N_c$ QCD \cite{Witten:1979kh}.
The quantum eigenstates have been obtained and interpreted as 
physical baryons in Ref.~\cite{Adkins:1983ya} and the corresponding
$S$-matrix describing meson-baryon scattering was computed in 
Refs.~\cite{Mattis:1984dh}-\cite{Karliner:1986wq} and in \cite{Walliser:wn}.

Still, one has to keep in mind that 
for QCD in 3+1 dimensions the exact low energy action is not known,
so one is forced to use symmetry properties and models to construct 
an approximate low energy effective action which admits soliton
solutions. The Skyrme model is just one specific representative 
of a large class of models of this kind.
The large-$N_c$ approximation is an important guiding principle
\cite{Witten:1979kh} for constructing such actions, but even after it is
implemented, much unknown remains.

In contrast, in 1+1 dimensions it is often possible to obtain analytic
solutions. Thus instead of studying an approximate
effective action in 3+1 dimensions, one can analyze the exact 
effective action in  1+1 dimensions. 
The limitations of such an approach are
obvious. The most striking one for QCD is the very different 
nature of chiral symmetry in 1+1 vs. 3+1 dimensions. 
Yet, for many purposes it is useful
to have a strong analytic grip on the 1+1 dimensional analogues of
the problems in 3+1 dimensions.

This line of thought has been pioneered by 't Hooft \cite{'tHooft:1974hx}
in his work on QCD$_2$ mesons in the large-$N_c$ limit. 
A lot of further progress in the study of QCD$_2$ was made possible 
through non-abelian bosonization and semiclassical methods applied to the
bosonized action.

Thus the baryon spectrum of QCD$_2$ for general $N_f$ and $N_c$ was
computed in Refs.~\cite{Date:1986xe},\cite{Frishman:1987cx};
the $\bar q q$ content of baryons was calculated in 
Ref.~\cite{Frishman:1990uw} and the physical picture of baryons composed
of constituent quark solitons was obtained in Ref.~\cite{Ellis:1992wu}.

In this paper we will compute the meson-baryon scattering in 
QCD$_2$ at strong coupling, following the techniques 
of \cite{Mattis:1984dh} 
applied to the bosonized action \cite{Frishman:1992mr}.

We first review the general method in Section 1, pointing out a
particularly simple way of parametrizing the small fluctuations around the
classical soliton.  

In Sec. 2 we then work out the abelian case of one flavor 
at strong coupling, as a ``warm-up".
It turns out that the resulting effective potential for small fluctuations
is of a special kind, with no reflection. Thus the modulus of the transmission
coefficient is 1, and the scattering is fully characterized by the phase
shift. We compute the phase shift, which changes smoothly from $\pi$ at the
threshold to 0 at infinite energy.

Motivated by the emergence of the reflectionless potential for this case,
we obtain in Sec. 3 the general condition for actions leading to
reflectionless effective potentials. In particular, we 
also find that  $(\phi^2-a^2)^2$ belongs to this category,
in addition to sine-Gordon action which appears in the strong-coupling
limit for one flavor.

Moving on to many flavors in Sec.~4, we find a relativistic analogue of a 
velocity-dependent potential, i.e.  an effective potential which
depends on the energy of the projectile.

In Sec.~5 we compute the energy dependence of the 
transmission and reflection amplitudes
as well as the phase shift. The latter is ${-}{\pi\over2}$ at the
threshold, and ${-}\pi$ at infinity, with a minimum at an energy 
of about 1.2 times the mass of the projectile. We find no evidence for
resonances.

\section{General Formulation} 

QCD in 1+1 dimensions, bosonized in the scheme
\beq
{[\,\,\underbrace{SU(N_c)\downstrut}\,\,]_{\lower0.5em\hbox{$N_F$}}\atop
\kern-1.4em h}
{\times\atop\phantom{A}}
{[\,\,\underbrace{U(N_F)\downstrut}\,\,]_{\lower0.5em\hbox{$N_c$}}\atop
\kern-1.4em g}
\eeq
is \cite{Frishman:1992mr}
\bea
S[g,h,A_+A_-] &=& N_c S[g] + N_f S[h]
\nonumber\\
\phantom{a}\nonumber\\
&+&{N_F\over 2\pi} \int d^2 x 
\,\hbox{Tr}_c \left[ i 
\left(
A_+ h \partial_- h^\dagger 
+
A_- h^\dagger \partial_+ h
\right)
-
\left(A_+ h A_- h^\dagger - A_- A_+ \right)\right]
\nonumber\\
\phantom{a}\nonumber\\
&-&{1\over 2 e_c^2}
\int d^2 x \,\hbox{Tr}_c \,F_{\mu\nu} F^{\mu\nu}
+ {m^\prime}^2 N_{\widetilde m} 
\int d^2 x \,\hbox{Tr} \left(g\,h + h^\dagger\,g^\dagger\right)
\\
\phantom{a}\nonumber\\
{m^\prime}^2 = m_q \,C\, \widetilde m &&
\nonumber
\eea
where $\widetilde m$ is to be fixed, $C=e^\gamma\approx 0.891$,
and
\bea
S[u] \equiv S_{WZW}[u] &=&
{1\over 8 \pi} \int d^2 x \,\hbox{Tr} 
\left(\partial_\mu u \partial^\mu u^\dagger\right)
\nonumber\\
\phantom{a}\nonumber\\
&+&{1\over 12 \pi} \int_B d^3 y \,  \epsilon_{ijk} \,\hbox{Tr} 
\left[
\left(u^{-1}\partial_i u \right)\,
\left(u^{-1}\partial_j u \right)\,
\left(u^{-1}\partial_k u \right)
\right]
\eea
In the strong coupling limit ${e_c/ m_q} \to \infty$
\cite{Frishman:1992mr}, 
\bea
\Seff &=& N_c S[g] + m^2 N_m \int d^2 x 
\left( \, \hbox{Tr}\, g + \, \hbox{Tr}\, g^\dagger \right)
\nonumber\\
\phantom{a}\nonumber\\
m &=& \left[ N_c C m_q \left({e_c \sqrt{N_f}\over
\sqrt{2\pi}}\right)^{\Delta_c}\right]^{1\over 1 + \Delta_c}
\\
\phantom{a}\nonumber\\
\Delta_c &=& { N_c^2 -1 \over N_c (N_c + N_F)}
\nonumber
\eea
Equation of motion, as coefficient of $(\delta g)g^\dagger$,
\beq
{N_c\over 4 \pi} \partial_+
\left[ \left(\partial_-g\right) g^\dagger\right]
+ m^2 \left( g - g^\dagger\right) = 0 
\eeq
We adopt the approach of Ref.~\cite{Mattis:1984dh} for deriving the 
meson-baryon scattering $S$-matrix, expanding in
small fluctuations around a given static classical solution
$g_c (x)=\exp[{-}i\Phi_c(x)]$,
\beq
g = \exp \left\{ 
{-}i 
\left[
\Phi_c(x) + \delta\phi(x,t)
\right]
\right\}
\eeq

\beq
g \approx
e^{ {-}i \Phi_c(x)}
-i
\int_0^1 d \tau\,
e^{ {-}i \tau \Phi_c(x)}
\delta\phi(x,t)
e^{ {-}i (1-\tau) \Phi_c(x)}
\eeq
%
%
\beq
\delta g = -i
\int_0^1 d \tau\,
e^{ {-}i \tau \Phi_c(x)}
\delta\phi(x,t)
e^{ {-}i (1-\tau) \Phi_c(x)}
\label{deltag}
\eeq
choose
\beq
\Phi_c(x) = 
\pmatrix{
\phi_c(x) &   &   &   &   &  \cr
          & 0 &   &   &   &  \cr
          &   & . &   &   &  \cr
          &   &   & . &   &  \cr
          &   &   &   & . &  \cr
          &   &   &   &   & 0 }
\eeq
Then
\beq
\phi_c(x)^{\prime\prime}-{8\pi\over N_c}m^2 \sin \phi_c = 0
\eeq
\beq
\partial_+
\left[
\left(
\partial_-\delta g \right) e^{i\phi_c(x)} 
-i \left(\partial_-\phi_c(x)\right)\,e^{{-}i\phi_c(x)} \delta g^\dagger
\right]
+m^2{4\pi\over N_c} \left( \delta g - \delta g^\dagger\right) = 0
\label{difeqI}
\eeq
\beq
\phi_c(x) = 4\hbox{arctg}\left(e^{\mu x}\right), \qquad  
\mu = m \sqrt{8 \pi \over N_c}
\eeq
Actually, to avoid integrals like in eq.~\eqref{deltag}, which yield rather
complicated expressions for the fluctuations via eq.~\eqref{difeqI}, we
will adopt a different expansion, namely
\bea
g&=&e^{{-}i \Phi_c(x)} \, e^{{-}i\tilde\delta \phi(x,t)}
\nonumber\\
\phantom{a}\\
&\approx&e^{{-}i\Phi_c(x)} - i e^{{-}i\Phi_c(x)}
\tilde\delta\phi(x,t)\nonumber
\label{newg}
\eea
where we have denoted by $\tilde\delta\phi$ the new variation, different
from the $\delta\phi$ of eq.~\eqref{deltag}, but still a fluctuation about
the classical solution.
Now
\beq
{N_c\over 4\pi} \,
\partial_+
\left[ 
e^{{-}i\Phi_c(x)} 
\left( \partial_-\tilde\delta\phi(x,t) \right)
e^{i\Phi_c(x)}
\right]
+ m^2 \,
\left[
e^{{-}i\Phi_c(x)}\tilde\delta\phi(x,t)+\tilde\delta\phi(x,t)
e^{i\Phi_c(x)}
\right]
= 0
\label{diffeqII}
\eeq
Obviously the two expressions coincide in the Abelian case.
In fact, the relation between $\delta \phi$ and $\tilde \delta\phi$ is
\beq
\tilde\delta \phi (x, t) = \int_0^1 d \tau \,e^{i \tau \Phi_c (x)}
\,\delta \phi (x, t) \,e^{{-}i \tau \Phi_c (x)}
\eeq
The results for physical quantities should be the same.

\section{Abelian Case}

As in the nonabelian case, we expand in small fluctuations around the
soliton \cite{Mattis:1984dh}.
Take $\delta\phi$ to commute with $\Phi_c$, in particular same entry in
matrix. Denote this case by $\delta\phi_A$, where the subscript ``A"
stands for ``Abelian".

Then
\beq
\delta g = {-} i \delta\phi_A(x) e^{{-}i\phi_c(x)}
\eeq
\beq
\Box \delta\phi_A + \mu^2(\cos\phi_c) \delta\phi_A = 0
\eeq
\beq
\cos\phi_c = \left[1 - {2\over \cosh^2 \mu x }\right]
\eeq
Get 
\bea
\Leff = {1\over2}\left(\partial_\mu \delta\phi_A\right)^2 
- {1\over2} V(x) \left(\delta\phi_A\right)^2\nonumber\\
\phantom{a}\\
V(x)=\mu^2 \cos\phi_c(x) = \mu^2 \left[1 - {2\over \cosh^2 \mu x }\right]
\label{Vofx}
\eea
Take
\beq
\delta\phi_A(x,t) = e^{{-}i\omega t} \chi_A(x)
\eeq
Then
\beq
{-}\omega^2 \chi_A - \chi_A^{\prime\prime} + V(x)\chi_A = 0
\label{potential}
\eeq

When $x\to \pm\infty$, the potential $\to\mu^2$, and so 
\beq
\chi_A^{\prime\prime}(\pm\infty) +
\omega^2\,\chi_A(\pm\infty) =
\mu^2\,\chi_A(\pm\infty)
\eeq
Take
\beq
\chi_A(x) \mathop{\longrightarrow}_{|x|\to\infty} e^{i k x}
\eeq
which results in
\beq
\omega^2 = k^2 +\mu^2
\eeq
We have two asymptotic solutions,
\beq
\chi_A(x) \sim A(\omega) \sin k x + B(\omega) \cos k x
\eeq
and the $S$ matrix is 
\bea
S_{\hbox{\small \it forward}} &=& {1\over2}(B - i A)
\nonumber\\
\phantom{a}\\
S_{\hbox{\small \it backward}}&=& {1\over2}(B + i A)
\eea
for an incoming wave $e^{i k x}$ from $x={-}\infty$.

We can now proceed to derive the scattering matrix, using the standard
procedure. The solution for
$x\to \infty$ contains only the transmitted wave,
$\psi (x\to \infty) \sim e^{i k x}$. 
\footnote{We take the convention where
the scattering phase is taken to be zero at $x\to \infty$ and is therefore
extracted from the wavefunction at $x\to{-}\infty$.}

\begin{figure}[h]
\bigskip
\centerline{\epsfig{file=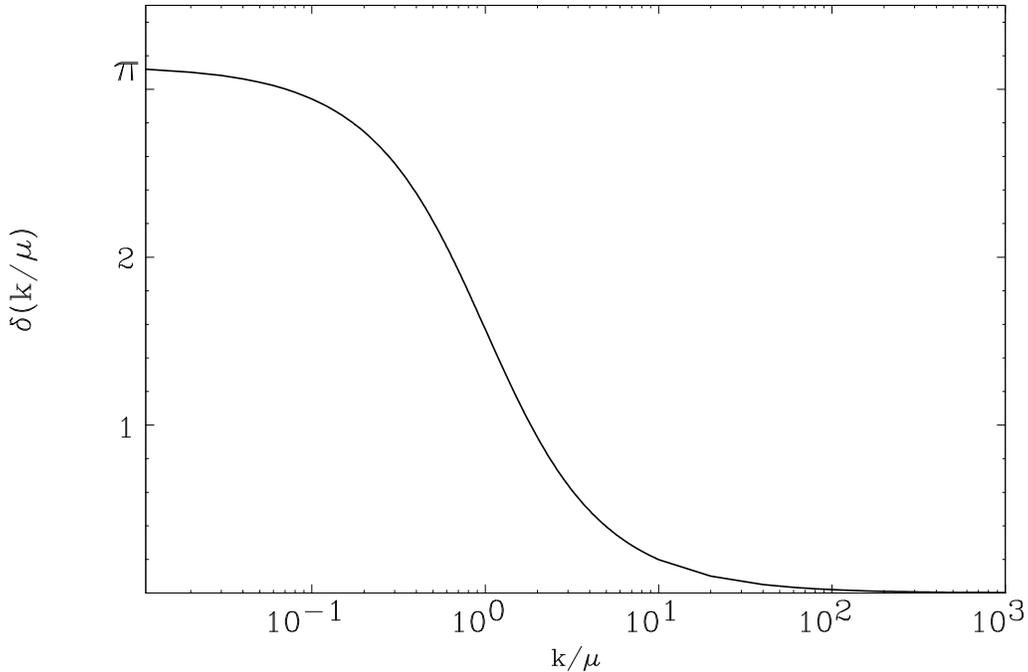,width=9cm,angle=90}}
\vspace{0.2in}
\caption{\small \it
The phase shift $\delta=2\hbox{ctg}^{-1}(k/\mu)$,
eq.~\protect\eqref{ctgdelta},
as function of the normalized momentum $k/\mu$,
 for the potential \eqref{Vofx}, governing the small
fluctuations around the soliton in the abelian case. 
The phase shift is smooth and monotonically decreasing with the momentum,
indicating that no resonance is present.
Note logarithmic momentum scale.}
\label{plotdelta_fig}
\end{figure}

It turns out that for the particular potential \eqref{Vofx} there is 
no reflection at all \cite{Landau}, i.e. the wavefunction for
$x\to{-}\infty$ contains only the incoming wave,
\beq
\psi(x\to{-}\infty) \sim e^{i k x-\delta} =
\left(
{-} {1+{i k /\mu} \over 1 - {i k /\mu}}
\right)
\,
 e^{i k x} 
\eeq
Thus
\beq
{1\over T} = {-} {1 + {i k/\mu} \over 1 - {i k/\mu}}
\eeq
\beq
T = {-} {1 - {i k /\mu} \over 1 + {i k/\mu}} = e^{i\delta}
\eeq
\beq
\hbox{ctg} \,{1\over2}\delta = {k\over \mu}
\label{ctgdelta}
\eeq
As shown in Fig.~\eqref{plotdelta_fig},
$\delta$ varies smoothly and decreases monotonically
from $\delta=\pi$ at $k=0$ to 
$\delta=0$ at $k=\infty$,
indicating that there is no resonance.

\section{No-Reflection Actions}

We got a no-reflection potential in the previous section, in the case
of one flavor. Reflectionless potentials are well-known in Quantum Mechanics
\cite{Landau}.
It is therefore interesting to find out about the general case 
in Quantum Field Theory
of an action which leads to a reflectionless potential 
of the form discussed in \cite{Landau}
 when one expands in
small fluctuations about the classical solution.

For other cases of reflectionless potentials see also
\cite{Novikov}.

Let us write
\beq 
{\cal L}={1\over2} \left(\partial_\mu\phi\right)^2 - W(\phi)
\label{genericL}
\eeq
where $W(\phi)$ is some potential.
Then
\beq
\Box \phi + W^\prime(\phi) = 0
\eeq
where $W^\prime(\phi)$ stands for $\delta W/\delta\phi$. Take a static
case, i.e.
\beq
\phi_s^{\prime\prime}(x) - W^\prime\left[\phi_s(x)\right] = 0
\eeq
where $\phi_s(x)$ is a classical, time-independent solution.  
Integrating
\beq
{1\over2}\left[\phi_s^\prime(x)\right]^2 = W\left[\phi_s(x)\right]
\eeq
where the constant of integration is zero, assuming that 
$\phi_s$ and $\phi_s^\prime$ are zero at $\pm \infty$ and that 
$W[0]=0$.

One more integration yields 
\beq
\int_0^{\phi_s(x)} {d \xi \over \sqrt{2 W(\xi)}} = x
\eeq
where we chose the origin so that there is no integration constant here as
well.

Let us now discuss small deviations from $\phi_s$,
\beq
\phi = \phi_s + \chi(x,t)
\eeq
Again
\beq
\chi(x,t)= e^{{-}i\omega t} \psi(x)
\eeq
Thus
\beq
{-}\psi^{\prime\prime}(x) + [W^{\prime\prime}(\phi_s) - \omega^2]
\psi(x) = 0
\eeq
Comparing with eq.~\eqref{potential}, we see that the potential is 
$W^{\prime\prime}\left(\phi_s(x)\right)$. Using
\beq
\phi_s^{\prime\prime\prime} = \phi_s^\prime {\delta^2
W(\phi_s)\over\delta\phi_s^2}
\label{Wdelta}
\eeq
we get 
\beq 
{\phi_s^{\prime\prime\prime}\over\phi_s^\prime}
=\mu^2 - { \tilde\mu^2 \,l (l+1)\over \cosh^2 \tilde\mu x}
\eeq
where $\mu$ is the mass associated with the fluctuations, namely
$\omega^2 = k^2 + \mu^2$, and $l$ is a positive integer.
Now, this potential has bound states at 
\cite{Landau}, \cite{Jaffe}
\beq
\omega_n^2 = \mu^2 -\tilde\mu^2\,(l-n)^2,
\qquad\qquad\qquad\qquad
n=0,1,\ldots,(l-1)
\eeq
corresponding to imaginary $k$, as expected. 

Now, eq.~\eqref{Wdelta} has a soliton for $\omega=0$, which is 
$\phi_s^\prime$, namely with fast decrease at $\pm\infty$, like a bound state.
We must thus have $\mu=\tilde\mu\, N$,  $1\le N\le l$.
Thus we get
\beq
{\phi_s^{\prime\prime\prime}\over\phi_s^\prime}
=
\tilde\mu^2
\left[ N^2 - { l (l+1)\over \cosh^2 \tilde\mu x} \right],
\qquad\qquad
\qquad\qquad
1\le N\le l
\eeq
The case $N=l=1$ can be integrated to give the sine-Gordon action,
with $\tilde\mu=m$, i.e. the mass in the original Lagrangian.

The case $l=N=2$ can be integrated too, giving
\beq
{\cal L} = {1\over2} \left(\partial_\mu\phi\right)^2 - {1\over4} m^2
\left(\phi^2 -a^2\right)^2
\eeq
this time with $\tilde\mu= m a/\sqrt{2}$. Note that now 
$\mu=\sqrt{2} m a$, which is also the mass of the shifted field
$\chi=\phi - a$.

We did not obtain explicit solutions for the case $N=l\ge 3$.
For implicit solutions, see \cite{Trullinger:ie}.

For the case $l=N+1$, we can integrate the equations for any $N$, finding that
\beq
\phi_s(x) = {C \over (\cosh \tilde\mu x )^N}
\eeq
and resulting in
\beq
W(\phi_s) = {1\over2} (\tilde\mu N)^2 
\left[ \phi_s^2 - C^{{-}2/N} \,\phi_s^{(2+{2\over N})} \right]
\eeq
But this yields an action in eq.~\eqref{genericL} that has no ground state, thus
not an interesting case.

\section{Non-Abelian Case}

Following eq.~\eqref{diffeqII}, we get
\beq
\Box \tilde \delta\phi - i \left( \partial_+\Phi_c \right)
\left( \partial_-\tilde\delta\phi \right)
+i\left(\partial_-\tilde\delta\phi\right)\left(\partial_+\Phi_c\right)
+\displaystyle{1\over2}\mu^2 
\left[
\tilde\delta\phi e^{-i\Phi_c(x)}
+e^{i\Phi_c(x)}\tilde\delta\phi
\right]
=0
\eeq
The equation for $\tilde\delta\phi_{i j}$  with ${i, j} \neq 1$ is like for 
the free case
\beq
\Box \tilde\delta \phi_{i j} + \mu^2 \tilde \delta\phi_{i j}  = 0\,,
\qquad\qquad i\ \hbox{and}\ j \neq 1
\eeq
whereas the $i=1,j=1$ matrix element is like in the abelian case
\beq
\Box \tilde\delta \phi_{11} + \mu^2 \left(\cos\phi_c(x)\right)
\tilde\delta \phi_{11} = 0
\eeq
with no reflection and no resonance.

So in order to proceed beyond these results, we need to consider
$\tilde\delta\phi_{1 j}$, $j\neq 1$,
or  $\tilde\delta\phi_{i 1}$, $i\neq 1$.
As $\tilde\delta\phi$ is hermitean, it is sufficient to discuss one of
the above.

\begin{figure}[h]
\bigskip
\centerline{\epsfig{file=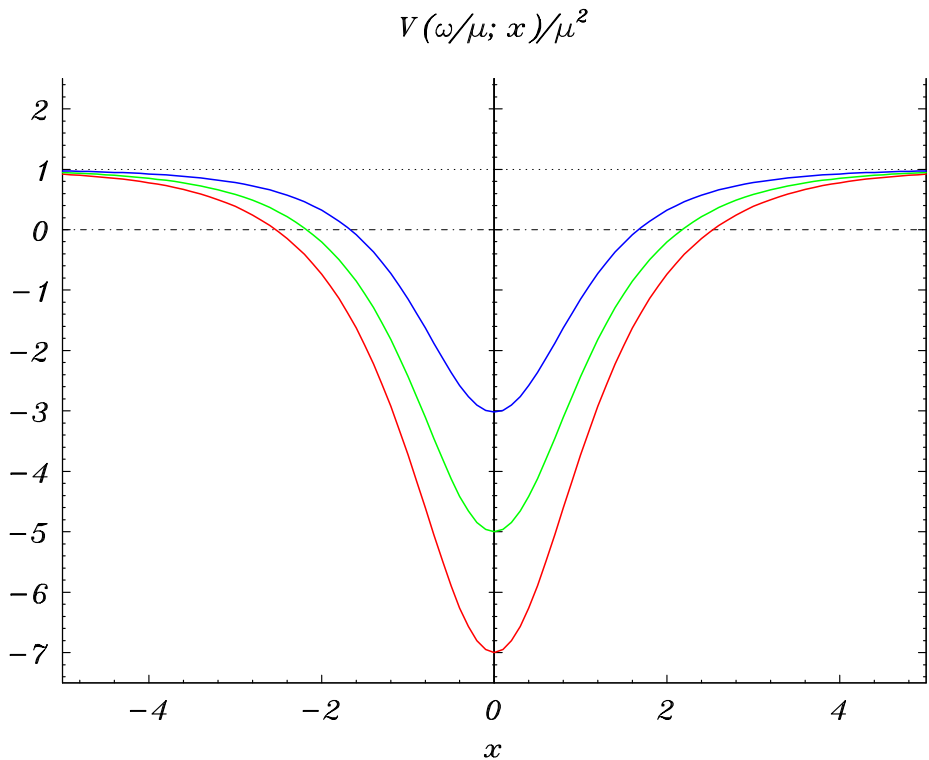,width=10.0cm,angle=90}}
\vspace{0.2in}
\caption{\small \it
The normalized potential $V(\omega/\mu;x)/\mu^2$ of 
eq.~\eqref{scatterV}, 
for $\omega/\mu=$1.01 (upper), 2 (middle) and 3 (lower).}
\label{scatterV_fig}
\end{figure}

Thus we take
\beq
\tilde\delta\phi_{1 j} =  e^{{-}i\omega t} u_j(x)
\qquad\qquad j\neq 1
\eeq
resulting in
\beq
u_j^{\prime\prime}(x) - i \phi_c^\prime(x) u_j^\prime(x)
+ \left[ \omega^2 + \omega \phi_c^\prime(x) - \textstyle {1\over2} \mu^2 
\left(1 + e^{i\phi_c(x)}\right)\right] u_j(x) = 0
\eeq
Define
\beq
u_j \equiv e^{{i\over2}\phi_c} \,v_j
\eeq
Then
\beq
v_j^{\prime\prime} + \left[ \omega^2 + \omega \phi_c^\prime 
- \textstyle{1\over2}
\mu^2\left(1+\cos\phi_c\right)
+\textstyle{1\over4}\left(\phi_c^\prime\right)^2
\right] v_j = 0
\label{DiffEqvj}
\eeq
Using
\beq
\textstyle{1\over2}\left(\phi_c^\prime\right)^2 =
\mu^2\left(1-\cos\phi_c\right)
\eeq
we get
\beq
v_j^{\prime\prime} + \left[ \omega^2 + \omega \phi_c^\prime
-\mu^2\cos\phi_c\right] v_j = 0 
\eeq
or
\beq
{-}v_j^{\prime\prime} - \omega^2 v_j + V(x) v_j = 0
\label{DiffEqvjI}
\eeq
where
\bea
V(x) &=& {-}\omega \phi_c^\prime + \mu^2\cos\phi_c =
\nonumber\\
\label{scatterV}
\phantom{a}\\
&=& \mu^2 - 2\mu^2 
\left[{(\omega/\mu)\over \cosh \mu x} + {1\over \cosh^2 \mu x}\right]
\nonumber
\eea
with $\omega=\sqrt{k^2+\mu^2}$ as before. 
Note that the potential depends on the momentum of the incoming particle, 
as shown in Fig.~\ref{scatterV_fig}.

\section{Numerical Results}

The usual textbook prescription for studying quantum mechanical
scattering in one dimension 
\cite{Landau} is to start at $t={-}\infty$ with an incoming wave 
$e^{i k x}$ at $x\sim {-}\infty$. One then analyzes the outgoing waves
at $t={+}\infty$, i.e. the transmitted wave
$T\,e^{i k x}$ for $x\sim {+}\infty$, and the reflected wave 
$R\,e^{-i k x}$ for $x\sim {-}\infty$. $T$ is the transition amplitude and 
$R$ is the reflection amplitude, with 
\beq
|T|^2 + |R|^2 =1
\eeq
ensured by unitarity. 

It turns out that for numerical solution of the
scattering problem it is more convenient to take the coefficient of
the outgoing wave at $x\sim {+}\infty$ to be 1,
instead of the $T$ prefactor, and
integrate eq.~\eqref{DiffEqvjI} backward, 
reading off the $T$ and $R$ amplitudes from the solution at 
$x\sim {-}\infty$.

\begin{figure}[!ht]
\centerline{\epsfig{file=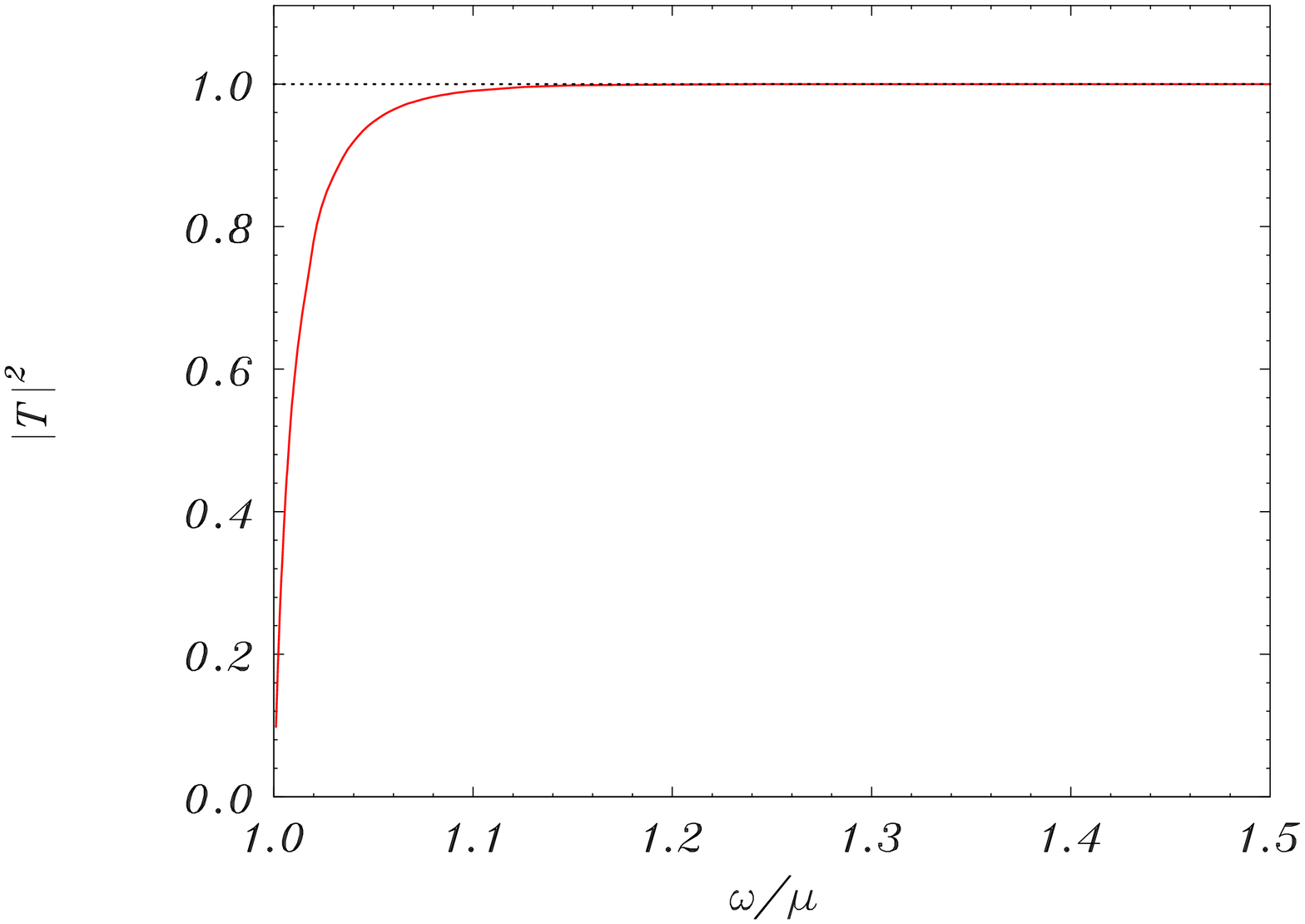,width=7.0cm,height=11cm,angle=90}}
\centerline{$\phantom{a}$}
\centerline{\epsfig{file=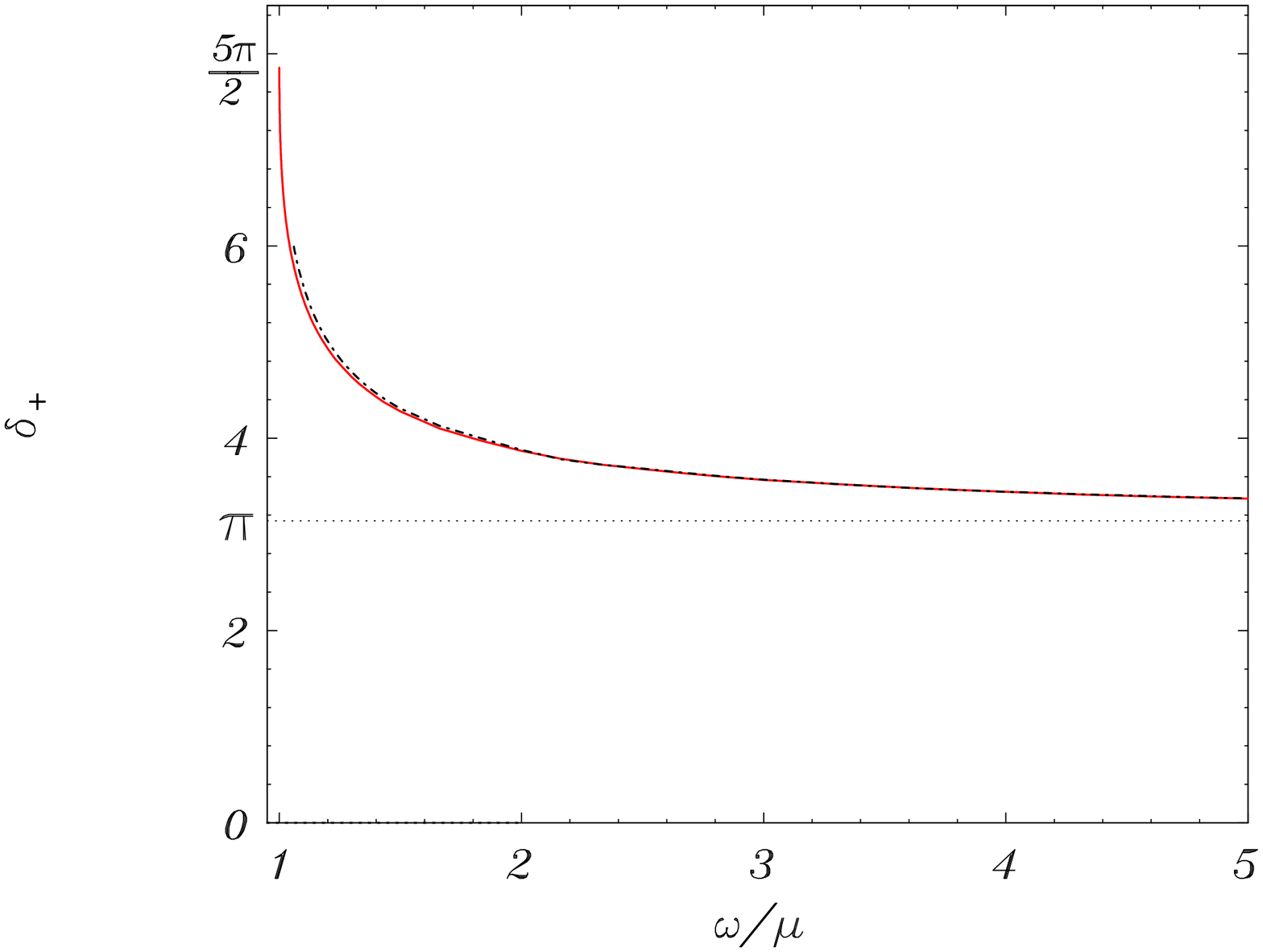,width=7.0cm,height=10.8cm,angle=90}}
\caption{\small \it
Scattering by the potential eq.~\eqref{scatterV} as function of the 
normalized energy $\omega/\mu$. Upper plot:
transmission probability $|T|^2$; \ lower plot:
phase of $T$, $\delta_+$ (continuous line). 
Also shown is the approximate 
result for $\delta_+$ from WKB (dot-dashed line).}
\label{T2delta_fig}
\end{figure}

We thus use
\bea
v_j(x) &=& e^{i k x},\qquad\qquad \phantom{aaaaaaaaaa}\,\,
x \,\to \,{+}\infty
\nonumber\\
\phantom{a}\\
v_j(x) &=& {1\over T} \, e^{i k x} + {R\over T}\, e^{-i k x},
\qquad\qquad x \,\to\, {-}\infty 
\nonumber
\eea
Since the potential is symmetric, the symmetric and anti-symmetric 
scattering amplitudes don't mix, yielding two independent phase shifts
$\delta_S$ and $\delta_A$, respectively. This leads to 
\bea
T&=&{1\over2} \left(e^{i\delta_S} + e^{i\delta_A}\right)
\nonumber\\
\phantom{a}\\
R&=&{1\over2} \left(e^{i\delta_S} - e^{i\delta_A}\right)
\nonumber
\eea
Define
\beq
\delta_{\pm}={1\over2}\left(\delta_S {\pm} \delta_A \right)
\label{deltaDef}
\eeq
\vbox{
\noindent
Then
\bea
T&=& e^{i\delta_+}\cos{\delta_-}
\nonumber\\
\phantom{a}\\
R&=&i e^{i\delta_+}\sin{\delta_-}
\nonumber
\eea }
Note that $R/T$ is purely imaginary.  
The transmission and reflections probabilities are
\bea
|T|^2 & = & \cos^2{\delta_-}
\nonumber\\
\phantom{a}\\
|R|^2  & = & \sin^2{\delta_-}
\nonumber
\eea
The numerical results for the transmission probability $|T|^2$
and for the phase of $T$, $\delta_+$ are presented in 
fig.~\ref{T2delta_fig}. 
For comparison and as an extra check we also plot the WKB 
result for $\delta_+$. 
Note that no resonance appears.

Note that the asymptotic value of the phase shift is $\pi$.
This can also be obtained from a WKB calculation,
which becomes exact at infinite energies.

\vfill\eject
 
\section*{Acknowledgments}
The research of one of us (M.K.) was supported in part by a grant from the
United States-Israel Binational Science Foundation (BSF), Jerusalem and 
by the Einstein Center for Theoretical Physics at the Weizmann Institute.

Y.F. would like to thank P. Breitenlohner and S. Ruijsenaars for discussions,
during a visit of his to the Max-Planck-Institut fuer Physik in Freimann. The 
hopitality of W. Zimmermann and the other members of the Institute is also
gratefully acknowledged.


\begin{thebibliography}{999}

\bibitem{Gervais:zg}
For reviews, see
{\em Extended Systems In Field Theory}, Proc.
Paris meeting, June 16-21, 1975,
J.L.~Gervais and A.~Neveu, Eds.,
Phys.\ Rept.\  {\bf 23} (1976) 237.

\bibitem{Rajaraman}
R. Rajaraman, {\em Solitons and Instantons:
an Introduction to Solitons and Instantons in Quantum Field Theory},
North-Holland, 2nd Ed., 1987.

\bibitem{Skyrme:vq}
T.~H.~Skyrme,
Proc.\ Roy.\ Soc.\ Lond.\ A {\bf 260}, 127 (1961).

\bibitem{Witten:1979kh}
E.~Witten,
Nucl.\ Phys.\ B {\bf 160}, 57 (1979).

\bibitem{Adkins:1983ya}
For an application of this procedure to the Skyrme model, see
G.~S.~Adkins, C.~R.~Nappi and E.~Witten,
Nucl.\ Phys.\ B {\bf 228}, 552 (1983).

\bibitem{Mattis:1984dh}
M.~P.~Mattis and M.~Karliner,
Phys.\ Rev.\ D {\bf 31}, 2833 (1985);
%
\bibitem{Karliner:1985ru}
M.~Karliner and M.~P.~Mattis,
Phys.\ Rev.\ Lett.\  {\bf 56}, 428 (1986).

\bibitem{Karliner:1986wq}
M.~Karliner and M.~P.~Mattis,
Phys.\ Rev.\ D {\bf 34}, 1991 (1986).

\bibitem{Walliser:wn}
H.~Walliser and G.~Eckart,
Nucl.\ Phys.\ A {\bf 429} (1984) 514.

\bibitem{'tHooft:1974hx}
G.~'t Hooft,
Nucl.\ Phys.\ B {\bf 75}, 461 (1974).

\bibitem{Date:1986xe}
G.~D.~Date, Y.~Frishman and J.~Sonnenschein,
Nucl.\ Phys.\ B {\bf 283}, 365 (1987);

\bibitem{Frishman:1987cx}
Y.~Frishman and J.~Sonnenschein,
Nucl.\ Phys.\ B {\bf 294}, 801 (1987).

\bibitem{Frishman:1990uw}
Y.~Frishman and M.~Karliner,
Nucl.\ Phys.\ B {\bf 344}, 393 (1990).

\bibitem{Ellis:1992wu}
J.~R.~Ellis, Y.~Frishman, A.~Hanany and M.~Karliner,
Nucl.\ Phys.\ B {\bf 382}, 189 (1992)
[arXiv:hep-ph/9204212].

\bibitem{Frishman:1992mr}
Y.~Frishman and J.~Sonnenschein,
Phys.\ Rept.\  {\bf 223}, 309 (1993)
[arXiv:hep-th/9207017].

\bibitem{Landau}
L.D. Landau and E.M. Lifshitz,
Quantum Mechanics: non-relativistic theory, 
3rd ed., \S23 and \S25, Pergamon Press, 1977.

\bibitem{Novikov} S.Novikov, S.V.Manakov, L.P.Pitaevskii and V.E.Zakharov,
{\em Theory of Solitons: The Inverse Scattering Method},
New York: Consultants Bureau, 1984.

\bibitem{Jaffe}
R.L. Jaffe, {\em An Algebraic Approach to Reflectionless Potentials in One
Dimension}, unpublished notes; available at\hfill\break
\def\mysim{\kern -.1667em\lower0.5ex\hbox{$\tilde{\phantom{a}}$}}
{\small \tt 
www-ctp.mit.edu/{\mysim}jaffe/8059\_98/Sample\_materials/SamplePaper.pdf}~.

\bibitem{Trullinger:ie}
S.~E.~Trullinger and R.~J.~Flesch,
J.\ Math.\ Phys.\  {\bf 28}, 1683 (1987).

\end{thebibliography}
\end{document}